\documentclass[epj,referee]{svjour}
\usepackage{indentfirst}
\usepackage{psfig,color}
\usepackage{epsfig}
\usepackage{epsf}
\usepackage{graphicx}

\renewcommand{\d}{{\mathrm d}}

\usepackage{graphics}
\begin{document}
\title{Collins Effect in Single Spin Asymmetries
of the $p^{\uparrow}p \to \pi X$ Process}

\author{Bo-Qiang Ma\inst{1} \and Ivan Schmidt\inst{2} \and Jian-Jun Yang\inst{3}\thanks{Deceased on June 11, 2004.} }

%
%
\institute{Department of Physics, Peking University, Beijing
100871, China \\ Di.S.T.A., Universit\`a del Piemonte Orientale
``A. Avogadro'' and INFN, Gruppo Collegato di Alessandria, 15100
Alessandria, Italy \and
Departamento de F\'\i sica, Universidad T\'ecnica
Federico Santa Mar\'\i a, Casilla 110-V, 
Valpara\'\i so, Chile \and Department of Physics, Nanjing Normal
University, Nanjing 210097, China \\ Departamento de F\'\i sica,
Universidad T\'ecnica
Federico Santa Mar\'\i a, Casilla 110-V, 
Valpara\'\i so, Chile\\
Institut f\"ur Theoretische Physik, Universit\"at Regensburg,
D-93040 Regensburg, Germany}

\date{Received: date / Revised version: date}
%
\abstract{We investigate the Collins effect in single spin
asymmetries (SSAs) of the $p^{\uparrow}p \to \pi X$ process, by
taking into account the transverse momentum dependence of the
microscopic sub-process cross sections, with the transverse
momentum in the Collins function integrated over. We find that the
asymmetries due to the Collins effect can only explain the
available data at best qualitatively, by using our choices of
quark distributions in the quark-diquark model and a pQCD-based
analysis, together with several options of the Collins function.
Our results indicate the necessity to take into account
contributions from other effects such as the Sivers effect or
twist-3 contributions.
} 

\maketitle

\begin{onecolumn}

Single spin asymmetries in hadronic reactions have become an
active research topic recently since they may help us to uncover
the transverse quark structure of the nucleon. Experimentally,
there are data on  meson electroproduction asymmetries in
semi-inclusive deep inelastic
scattering~\cite{HERMES00,SMC99,HERMES03}, whose most common
explanation is to relate them to the transversity distribution of
the quarks in the hadron ~\cite{Jaff96,Boer02}, convoluted with a
transverse momentum dependent fragmentation
function~\cite{Col93,Barone02,MSY9,MSY10-13,Efr}, i.e., the
Collins function~\cite{Col93,Kot95}. The Collins function, which
gives the distributions for a transversely polarized quark to
fragment into an unpolarized hadron with non-zero transverse
momentum, has aroused great interest recently since a chiral-odd
structure function can be accessible together with another
chiral-odd distribution/fragmentation function.

Large single spin asymmetries (SSA) in $p^{\uparrow}p \to \pi X$
have been also observed by the E704 Group at Fermilab~\cite{E704}.
Anselmino, Boglione and Murgia~\cite{Anselmino99} tried to
reproduce the experimental data using a parametrization of the
Collins fragmentation function, with several  assumptions and in a
generalized factorization scheme at the parton level. Notice that
the Collins function obtained in Ref.~\cite{Anselmino99} is
different from the original Collins parametrization~\cite{Col93}.
Other mechanisms such as the Sivers
effect~\cite{Sivers,BHS1,BHS2,sivers2}, and twist-3
contributions~\cite{Qiu98,KK00,3twist}, have been also proposed.
Very recently, Bourrely and Soffer~\cite{BS03} pointed out that
the SSA observed several years ago at FNAL by the experiment
E704~\cite{E704} and the recent result observed at BNL-RHIC by
STAR~\cite{STAR03} are due to different phenomena. It is thus
important to check whether a calculation with more detailed
microscopic dynamics taken into account, and with also proper
constraints on the Collins function, can indeed explain the single
spin asymmetries in $p^{\uparrow}p \to \pi X$. We will show in the
following that the Collins effect in single spin asymmetries
(SSAs) of the $p^{\uparrow}p \to \pi X$ process can only explain
the available experimental data at best qualitatively, but we have
not been able to explain the magnitude of the data at large $x_F$
by using several options of the Collins function and with our
choices of the quark distributions in the quark-diquark model and
a pQCD-based analysis. In our analysis, we work within the same
simplified planar configuration of Ref.~\cite{Anselmino99}, using
several options of Collins function with a Gaussian-type
transverse momentum dependence, together with our choices of quark
distributions. Notice also that measurements of SSAs in weak
interaction processes can distinguish between different QCD
mechanisms~\cite{BHS2}.

We first focus on the formalism of description of single spin
asymmetries at the parton level. For the inclusive production of a
hadron $C$ from the hadron $A$ and hadron $B$ collision process
\begin{equation}
A+B \to C+X,
\end{equation}
the Mandelstam variables $s$, $t$, and $u$ are written as
\begin{eqnarray}
&s=(P_A+P_B)^2=M_A^2+M_B^2+ 2 P_A \cdot P_B,\\
&t=(P_A-P_C)^2=M_A^2+M_C^2-2 P_A \cdot P_C,\\
&u=(P_B-P_C)^2=M_B^2+M_C^2-2 P_B \cdot P_C.
\end{eqnarray}
The experimental cross sections are usually expressed in terms of
the experimental variables $x_F=2P_L/\sqrt{s}$ and $P_T$ at a
given $s$, where $P_L=x_F \sqrt{s}/2$ and $P_T$ are the
longitudinal and transverse momenta of the produced hadron $C$
respectively, with energy $E_C =\sqrt{M_C^2+P_C^2}=
\sqrt{M_C^2+P_L^2+P_T^2}=\sqrt{M_C^2+P_T^2+x_F^2 s/4}$ in the
center of mass frame of the collision process, and $s$ is the
squared center of mass energy. The momentum and the energy of
hadron $A$ in the center of mass frame are
\begin{equation}
P_{AL}=\left| \mathbf{P}_A\right|=\sqrt{\frac{(s-M_A^2-M_B^2)^2-4
M_A^2 M_B^2}{4s}},
\end{equation}
\begin{equation}
E_A=\sqrt{M_A^2+{P^2_{AL}}},
\end{equation}
and the momentum and the energy of hadron $B$ are
\begin{equation}
\mathbf{P}_B=-\mathbf{P}_A,
\end{equation}
\begin{equation}
E_B=\sqrt{M_B^2+{P^2_{AL}}}.
\end{equation}
The Mandelstam variables $t$ and $u$ can be expressed as
\begin{eqnarray}
&t=(P_A-P_C)^2=M_A^2+M_C^2-2 E_A  E_C+2 {P_{AL}} P_L,\\
&u=(P_B-P_C)^2=M_B^2+M_C^2-2 E_B  E_C-2 {P_{AL}} P_L.
\end{eqnarray}
If the transverse momentum of the produced hadron relative to the
fragmenting quark is negligible and the factorization theorem
applies, the cross section can be written in terms of the
subprocess at the parton level as
\begin{equation}
\d \sigma=\frac{E_C \d ^3 \sigma^{AB \to C X}}{\d ^3 {\mathbf
P}_C} =\sum_{a,b,c,d}\int \frac{\d x_a \d x_b \d z }{\pi
z^2}f_{a/A}(x_a) f_{b/B}(x_b) \hat{s}
\delta(\hat{s}+\hat{t}+\hat{u})\frac{\d \hat{\sigma}^{a b \to c d
}}{\d \hat{t} }(x_a,x_b,z)D_{C/c}(z).
\end{equation}
However, there is no such a theorem if one considers transverse
parton momenta as done sequently by taking the Collins effect into
account. The Collins effect consists in taking into account the
transverse momentum $\mathbf{k}_{c\perp}$ of the produced hadron
$C$ relative to the fragmenting parton $c$ with momentum ${\mathbf
p}_c$, so we have
\begin{equation}
{\mathbf P}_C=z {\mathbf p}_c+ {\mathbf k}_{c\perp},
\end{equation}
where $z$ is the momentum fraction of the produced hadron relative
to the fragmenting parton. The cross section is then expressed as
\begin{equation}
\begin{array}{lll}
\d \sigma=\frac{E_C \d ^3 \sigma^{A B \to CX}}{\d ^3 {\mathbf
P}_C}
\\ \nonumber
=\sum_{a,b,c,d}\int  \d^2 {\mathbf k}_{c\perp} \frac{\d x_a \d x_b
\d z }{\pi z^2} f_{a/A}(x_a) f_{b/B}(x_b) \hat{s}
\delta(\hat{s}+\hat{t}+\hat{u})\frac{\d \hat{\sigma}^{a b \to c d
}}{\d \hat{t} }(x_a,x_b,z,{\mathbf k}_{c\perp})D_{C/c}(z,{\mathbf
k}_{c\perp}),
\end{array}
\end{equation}
where $D_{C/c}(z,{\mathbf k}_{c\perp})$ describes the
fragmentation of hadron $C$ with longitudinal momentum fraction
$z$ and transverse momentum ${\mathbf k}_{c\perp}$ relative to the
fragmenting parton $c$. The denotations $a$, $b$, and $c$ should
include both the flavors and respective polarizations of the
involved partons, if any or some of the incident hadrons $A$, $B$,
and the produced hadron $C$ are polarized. The cross section of
the subprocess $a +b \to c +d$
\begin{equation}
\d \hat{\sigma}=\frac{\d \hat{\sigma}^{a b \to c d }}{\d \hat{t}
}(x_a,x_b,z,{\mathbf k}_{c\perp}),
\end{equation}
should be written in terms of the Mandelstam variables $\hat{s}$,
$\hat{t}$, and $\hat{u}$
\begin{eqnarray}
&\hat{s}=2 p_a \cdot p_b=x_a x_b s,\\
&\hat{t}=-2 p_a \cdot p_c=\frac{x_a}{z} t \Phi_t(\pm k_{c\perp}),\\
&\hat{u}=-2 p_b \cdot p_c =\frac{x_b}{z} u \Phi_u(\pm k_{c\perp}),
\end{eqnarray}
where the functions $\Phi_t(\pm k_{c\perp})$ and $\Phi_u(\pm
k_{c\perp})$ are given by~\cite{ABDM01}
\begin{equation}
\Phi_t(\pm k_{c\perp})=g(k_{c\perp})\left\{1 \mp 2
k_{c\perp}\frac{\sqrt{stu}}{t(t+u)}
-\left[1-g(k_{c\perp})\right]\frac{t-u}{2 t}\right\},
\end{equation}
\begin{equation}
\Phi_u(\pm k_{c\perp})=g(k_{c\perp})\left\{1 \pm 2
k_{c\perp}\frac{\sqrt{stu}}{u(t+u)}
+\left[1-g(k_{c\perp})\right]\frac{t-u}{2 u}\right\},
\end{equation}
with $g(k_{c\perp})=\sqrt{1-k_{c \perp}^2/P_C^2}$, and $\pm
k_{c\perp}$ referring respectively to the configuration in which
${\mathbf k}_{c\perp}$ points to the left or to the right of
${\mathbf p}_c$. The kinematical effect from the transverse
momentum ${\mathbf k}_{c\perp}$ is explicitly taken into account
in $\Phi_t(\pm k_{c\perp})$ and $\Phi_u(\pm k_{c\perp})$, which
become $1$ for $k_{c\perp}=0$. The four variables $x_a$, $x_b$,
$z$, and $k_{c\perp}$ are not independent, and by exploiting the
$\delta(\hat{s}+\hat{t}+\hat{u})$ function we get
\begin{equation}
x_a x_b s+\frac{1}{z}\left[x_a t \Phi_t(\pm k_{c\perp}) +x_b u
\Phi_u(\pm k_{c\perp}) \right]=0,
\end{equation}
so we have
\begin{equation}
z=-\frac{x_a t \Phi_t(\pm k_{c\perp}) +x_b u \Phi_u(\pm
k_{c\perp}) }{x_a x_b s}.
\end{equation}
After integrating the $\delta$-function, we get the cross section
\begin{equation}
\begin{array}{lll}
\d \sigma&=&\frac{E_C \d ^3 \sigma^{A B \to CX}}{\d ^3 {\mathbf
P}_C}
\\
\nonumber &=&\sum_{a,b,c,d}\int \d^2 {\mathbf k}_{c\perp} \left\{
\int_{x_{a}^{min}}^1 \d x_a \int_{x_{b}^{min}}^1 \d x_b
f_{a/A}(x_a) f_{b/B}(x_b) \frac{1}{\pi z} \frac{\d \hat{\sigma}^{a
b \to c d }}{\d \hat{t} }(x_a,x_b,z,{\mathbf k}_{c\perp})\right\}
D_{C/c}(z,{\mathbf k}_{c\perp}),
\end{array}
\label{scC}
\end{equation}
where
\begin{equation}
x_{a}^{min}=-\frac{x_b u \Phi_u(\pm k_{c\perp})}{x_b s+t
\Phi_t(\pm k_{c\perp})}, \ \ x_{b}^{min}=-\frac{t \Phi_t(\pm
k_{c\perp})}{ s+u \Phi_u(\pm k_{c\perp})}.
\end{equation}
Because of the $k_{c\perp}$ dependence of $x_{a}^{min}$ and
$x_{b}^{min}$, the integration over ${\mathbf k}_{c\perp}$ should
be performed after the integrations over $x_a$ and $x_b$, and $
{\mathbf k}_{c\perp}$ should meet the constraint $ k_{c\perp}\le
P_C$. Once we know the quark distributions $f_{a/A}(x)$ and
$f_{b/B}(x)$, the cross sections $\d \hat{\sigma}=\frac{\d
\hat{\sigma}^{a b \to c d }}{\d \hat{t}
}(\hat{s},\hat{t},\hat{u})$, and the fragmentation functions
$D_{C/c}(z,{\mathbf k}_{c\perp})$, we can calculate the cross
section (\ref{scC}) explicitly.

Now we apply the above formulas to the single spin asymmetries
\begin{equation}
A(x_F,P_T)=\frac{\d \sigma^{\uparrow}-\d \sigma^{\downarrow}}{\d
\sigma^{\uparrow}+\d \sigma^{\downarrow}}
\end{equation}
of $p^{\uparrow}p \to \pi X$ process, measured by the E704 Group
at Fermilab~\cite{E704}. After a calculation of the cross section
in terms of the explicit polarization dependent ingredients, we
find that the above asymmetries can be written as
\begin{equation}
A(x_F,P_T)=\frac{\int \d^2 {\mathbf k}_{c\perp} \d x_a \d x_b
\delta q_a(x_a) q_b(x_b) \Delta
\hat{\sigma}(\hat{s},\hat{t},\hat{u}) \Delta D^N(z,{\mathbf
k}_{c\perp})} {\int \d^2 {\mathbf k}_{c\perp} \d x_a \d x_b
q_a(x_a) q_b(x_b) \hat{\sigma}(\hat{s},\hat{t},\hat{u})
D(z,{\mathbf k}_{c\perp})}, \label{SSApi}
\end{equation}
where $\delta q_a(x_a)$ is the quark transversity distribution,
$q_b(x_b)$ is the usual quark distribution, $ \d \hat{\sigma}=\d
\hat{\sigma}^{\uparrow\uparrow}+\d
\hat{\sigma}^{\uparrow\downarrow}$, $\Delta \d \hat{\sigma}=\d
\hat{\sigma}^{\uparrow\uparrow}-\d
\hat{\sigma}^{\uparrow\downarrow}$, $D(z,{\mathbf
k}_{c\perp})=[D_{\pi/c^{\uparrow}}(z,{\mathbf
k}_{c\perp})+D_{\pi/c^{\downarrow}}(z,{\mathbf k}_{c\perp})]/2$,
and $\Delta D^N(z,{\mathbf
k}_{c\perp})=D_{\pi/c^{\uparrow}}(z,{\mathbf
k}_{c\perp})-D_{\pi/c^{\downarrow}}(z,{\mathbf k}_{c\perp})$ is
the Collins function.

For the quark distributions, we use those in a quark-diquark
model~\cite{Ma96} and a pQCD based analysis~\cite{Bro95},
explicitly taking into account the sea contributions based on the
GRV parametrization of the parton distribution functions in
Ref.~\cite{GRV95}. The unpolarized and transversely polarized
cross sections of the subprocess at the parton level,
$\hat{\sigma}(\hat{s},\hat{t},\hat{u})$ and $\Delta
\hat{\sigma}(\hat{s},\hat{t},\hat{u})$, can be found in
Refs.~\cite{BRST,SV}. The detailed information on the quark
transversity distributions in the quark-diquark model and the pQCD
based analysis can be found in Ref.~\cite{MSY10-13}. We need to
point it out here that these transversity distributions do not
come from model calculations, but from relations that connect the
transversity distributions with parametrized unpolarized quark
distributions, so that the calculation can be performed at the
same scale at the experiment. It is important to use a same set of
both unpolarized and polarized quark distributions, otherwise it
is not reasonable to compare the denominator with the numerator.
The calculated ratio would be unreasonable if the transversity
distributions were taken from a model and then use parton
distributions from a parametrization in order to perform the
calculation. This aspect has been carefully considered in our
calculation. In a strictly sense, the unpolarized quark
distributions and transversity distributions evolve differently,
so that we should use the relations at a specific initial scale
such as $Q^2 \approx 2~\mbox{GeV}^2$, and then consider the
evolution of the numerator and denominator separately. However,
the effects of evolution are presumably smaller than other
uncertainties in the approach, such as the neglect of unfavored
fragmentation, so we neglect the explicitly difference in
evolution and take the energy scale the same in both the numerator
and denominator. Besides, the quark-diquark model~\cite{Ma96} has
been successful in providing good descriptions of the quark
helicity distributions from experiments~\cite{JLab}, as well as
data for nucleon form factors~\cite{MQS}, so the extension of
applying it to the transversity distributions is reasonable.

For the pion fragmentation functions, we take only the favored
fragmentation into account, following
Ref.~\cite{Anselmino99}, and adopt the Kretzer-Leader-Christova
parametrization~\cite{Kre01} of $D(z)$
\begin{equation}
\begin{array}{ll}
D(z)=0.689 z^{-1.039}(1-z)^{1.241}.
\label{DDf}
\end{array}
\end{equation}
The scale for $D(z)$ is $\left<Q^2\right> = 2.5~\mbox{GeV}^2$. In
our case $s = 400~\mbox{GeV}^2$, but what matters for the
evolutions is $P_L^2$. Now $P_L = x_F \sqrt{s}/2$, which is about
$5~\mbox{GeV}$, and then $P_L^2$ is about $25~\mbox{GeV}^2$. We
neglect the difference between different fragmentation functions
in the evolution, as the effect due to the evolution, which is at
most logarithmic, can be reasonably neglected in predicting the
asymmetries when only ratios between different fragmentation
functions are relevant. To take into account the ${\mathbf
k}_{c\perp}$-dependence, we parameterize the fragmentation
function $D(z,{\mathbf k}_{c\perp})$ as~\cite{MSY10-13}
\begin{equation}
D(z,k_{c\perp})=\frac{R^2}{\pi} D(z) \exp(-R^2 k_{c\perp}^2),
\end{equation}
where $R=\frac{1}{\langle k_{c\perp}^2 \rangle ^{1/2}}$ with
$\langle k_{c\perp}^2 \rangle ^{1/2}=0.44 ~ \rm{GeV}$. The Collins
function $\Delta D (z,k_{c\perp})$ is ${\mathbf k}_{c\perp}$ odd,
and it gives a null contribution in case $k_{c\perp}=0$, so we
adopt the Collins parametrization~\cite{Col93} for $\Delta D
(z,k_{c\perp})$ as follows:
\begin{equation}
\frac{\Delta D (z,k_{c\perp})}{2 D (z,k_{c\perp})}= \frac{2 M_C
(k_{c\perp}/z)}{(M_C^2 +k_{c\perp}^2/z^2)}, \label{CollinsPara}
\end{equation}
where $M_C$ is a typical hadronic scale around $0.3 \to 1~
\rm{GeV}$ and we take its value as $0.7~ \rm{GeV}$ in our
calculation. The above Collins parametrization fulfills the
bound~\cite{Anselmino99}:
\begin{equation}
\frac{\left|\Delta D (z,k_{c\perp})\right|}{2 D (z,k_{c\perp})}
\le 1 .
\end{equation}

In the above formulas, all quantities needed in the calculation
are clearly given except that there are still large uncertainties
concerning the Collins function $\Delta D (z,k_{c\perp})$.
Therefore we consider several options for the Collins function:

\begin{enumerate}

\item

Option 1: The Collins parametrization  Eq.~(\ref{CollinsPara}).

\item

Option 2: The Collins parametrization with an additional
$z$-dependent factor
\begin{equation}
\frac{\Delta D (z,k_{c\perp})}{2 D (z,k_{c\perp})}= \frac{2 M_C
(k_{c\perp}/z)}{(M_C^2 +k_{c\perp}^2/z^2)}P(z), \label{Opt2}
\end{equation}
where $P(z)=c z^\alpha (1-z)^\beta$ with $\alpha=15$, $\beta=1.5$,
and $c=306.18$. This option is chosen to illustrate a case which
is of similar shape and not far from that of the
Anselmino-Boglione-Murgia (ABM) parametrization (Option 3), but
which can reproduce the asymmetries with magnitude around twice
that of Option 3.

\item

Option 3: The Anselmino-Boglione-Murgia (ABM) parametrization
(there is some trivial difference in details)~\cite{Anselmino99}
\begin{equation}
\frac{\Delta D (z,k_{c\perp})}{2 D (z,k_{c\perp})}= \left\{
\begin{array}{lll}
& 0.007409z(1-z)^{-1.3}\ \ &z < 0.97742;\\
&1 \ \ &z \ge 0.97742,
\end{array}
\right. \label{Opt3}
\end{equation}
at an average value of $\left< k_{c \perp}^2\right>^{1/2}=k_{c
\perp}^0$.

\item

Option 4: The upper limit bound
\begin{equation}
\frac{\Delta D (z,k_{c\perp})}{2 D (z,k_{c\perp})}=1. \label{Opt4}
\end{equation}

\end{enumerate}

The $z$-dependence of the ratio $\frac{\Delta D (z,k_{c\perp})}{2
D (z,k_{c\perp})}$  at a given $k_{c\perp}=k_{c\perp}^0=M_C$ for
the four options can be found in Fig.~\ref{msy14f1}, from where we
can find that the shapes of the Collins function are quite
different for the four options.

\vspace{0.3cm}
\begin{figure}[htb]
\begin{center}
\leavevmode {\epsfysize=5.5cm \epsffile{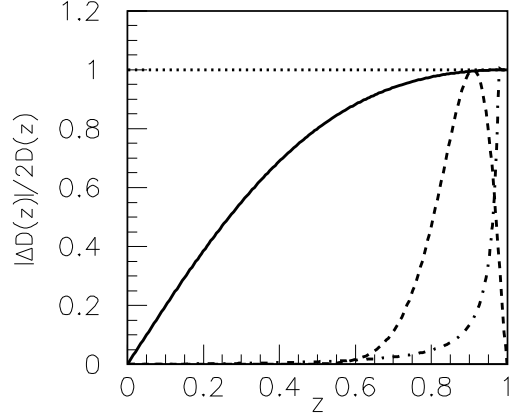}}
\end{center}
\caption[*]{\baselineskip 13pt The $z$-dependence the ratio
$\frac{\Delta D (z,k_{c\perp})}{2 D (z,k_{c\perp})}$ at a given
value of $k_{c\perp}=k_{c\perp}^0=M_C$ for the four options of the
Collins function. The solid, dashed, dash-dotted, and dotted
curves are corresponding to the results for the 4 options of
Collins function respectively. }\label{msy14f1}
\end{figure}

Thus we have all of the ingredients for the calculation of the
single spin asymmetries $A(x_F,P_T)$ in (\ref{SSApi}). The results
shown in Fig.~\ref{msy14f2} are calculated at $P_T=1.5~\mbox{GeV}$
with the quark transversity distributions from both the
quark-diquark model and the pQCD based analysis. We notice that
both models give large $A(x_F,P_T)$ at small $x_F$, with magnitude
compatible with or larger than the data, but at large $x_F$ the
magnitude is  below the data.   The quark-diquark model can
reproduce the trend of $A(x_F,P_T)$ for $\pi^-$ qualitatively, but
the pQCD based analysis produces a result with opposite sign. The
reason for the discrepancy is that the valence $d$-quark
transversity distribution is positive at large $x$ in the pQCD
based analysis~\cite{MSY10-13}. The situation for the
quark-diquark model is better, but there is still a discrepancy
with the magnitude of the data. In order to reduce the discrepancy
between the calculated result and the data, especially for the
$\pi^-$ data, we introduce an additional case (which is not a
realistic case, because it breaks the Soffer's
inequality~\cite{Soffer}, so this is just as an illustration) in
the quark-diquark model with the valence $d$-quark transversity
distributions enhanced by a factor of 3, i.e.,
\begin{equation}
\delta d_v(x)=-d_v(x) \hat{W}_v(x),
\end{equation}
where $d_v(x)$ is the unpolarized quark distribution and
$\hat{W}_v(x)$ is a Melosh-Wigner rotation factor to reflect the
relativistic effect from quark transverse motions~\cite{Ma91}. We
find that the magnitude of the calculated results is more
compatible with the data than both the quark-diquark model and the
pQCD based analysis, as shown in Fig.~\ref{msy14f2}. This might
suggest that the $d$-quark transversity distributions is more
negatively polarized than predicted in the quark-diquark model, as
supported by sum-rule based arguments~\cite{Ma98b}, if one only
considers the Collins effect. From the results of the upper limit
bound of the Collins function, we notice that any improvement in
the parametrization of the Collins function cannot improve the fit
to the data at large $x_F$.  Inclusion of unfavored
fragmentation~\cite{MSY10-13} will lead the calculated results to
go in the opposite direction from the data. The ABM
parametrization reproduces the shape of the data, but
underestimates its magnitude by a factor of around 4-6. We have
neglected the gluon transverse polarization
in our calculation, as the gluon transversity in a spin 1/2 hadron
is strictly zero due to helicity conservation.
Thus the description of the data
is not obtained in our calculation by taking into account the
detailed microscopic cross sections with the transverse momentum
in the Collins function integrated over, together with our choices
of quark distributions and Collins function.

\vspace{0.3cm}
\begin{figure}
\begin{center}
\leavevmode {\epsfysize=11.5cm \epsffile{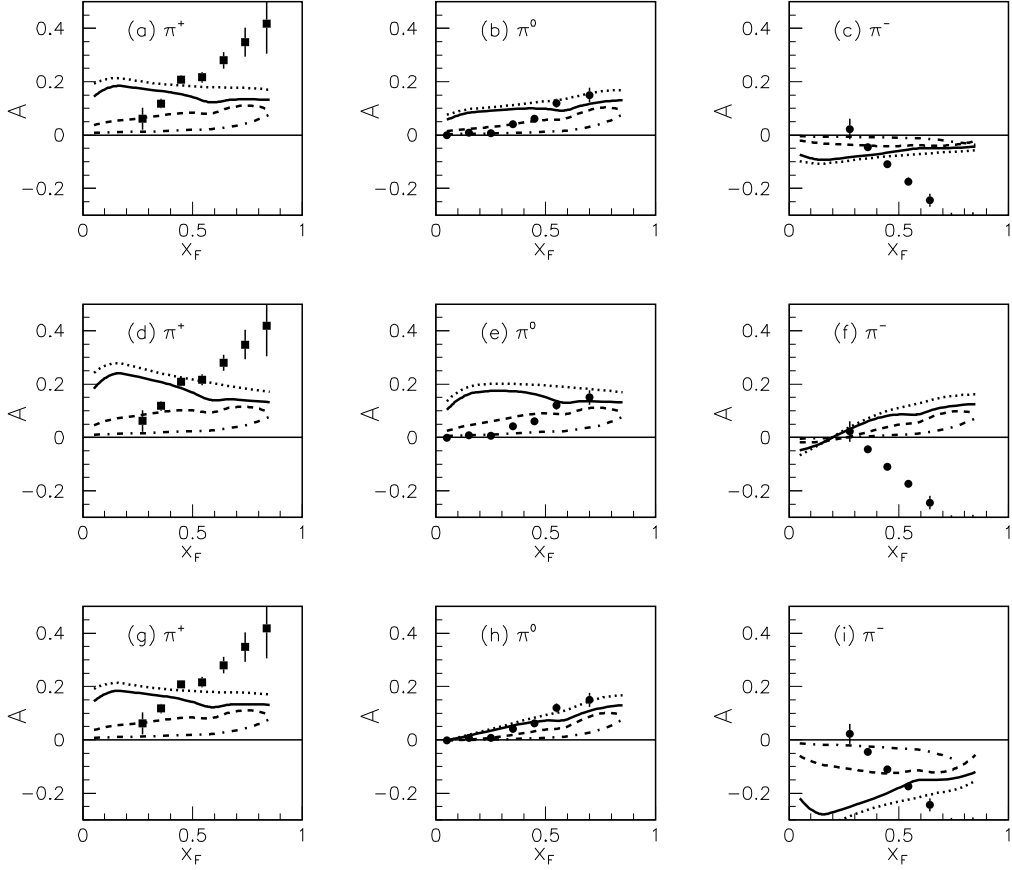}}
\end{center}
\caption[*]{\baselineskip 13pt The single spin asymmetries of
$p^{\uparrow}p \to \pi X$ process. The calculated results in the
quark-diquark model: (a), (b), and (c), in the pQCD based
analysis: (d), (e), and (f), and in the quark-diquark model with
more negatively polarized valence $d$-quark: (g), (h), and (i).
The solid, dashed, dash-dotted, and dotted curves are
corresponding to the results for the 4 options of Collins function
respectively. The experimental data are given by the E704
Group~\cite{E704} }\label{msy14f2}
\end{figure}

In summary, we checked the Collins effect in single spin
asymmetries (SSAs) of the $p^{\uparrow}p \to \pi X$ process  by
taking into account the transverse momentum dependence of the
microscopic sub-process cross sections at the parton level, with
the transverse momentum in the Collins function integrated over.
We introduced several options for the Collins effect, and found
that the single spin asymmetries of $p^{\uparrow}p \to \pi X$
process due to Collins effect can only explain the available data
at best qualitatively in some specific situations, by using the
quark distributions in the quark-diquark model and in a pQCD-based
analysis. The results suggest the necessity of taking into account
contributions from other effects such as Sivers effect and twist-3
contributions. It might be also possible that some unexpected
novel behaviors of the quark distributions and Collins function
need to be introduced. With our present calculation results on the
asymmetries and  the calculation done in Ref.~\cite{BS03} for the
cross sections themselves, we have to conclude that one needs to
introduce other mechanisms in order to understand  the single-spin
asymmetries for $\pi$ inclusive production in $pp$ collisions.  A
similar conclusion with a more detailed and complete analysis has
been also drawn in a recent work~\cite{Aetal}.

{\bf Acknowledgments: } We acknowledge the helpful comments and
suggestions from Andreas Sch\"afer and Mauro Anselmino. This work
is partially supported by National Natural Science Foundation of
China, by the Key Grant Project of Chinese Ministry of Education
(NO.~305001), by Fondecyt (Chile) grant 1030355, by Alexander von
Humboldt-Stiftung (J. J. Yang), and by the Italian Ministry of
Education, University and Research (MIUR).

\end{onecolumn}


\end{document}